\documentclass[12pt]{article}
\usepackage{epsfig}

\newcommand{\Li}{{\rm Li}}
\newcommand{\msbar}{\overline{\rm MS}}
\newcommand{\pfrac}[2]{\left(\frac{#1}{#2}\right)}
\newcommand{\slq}{q\kern-5.5pt/}
\newcommand{\slv}{v\kern-5pt\raise1pt\hbox{$\scriptstyle/$}\kern1pt}
\begin{document}
\begin{flushright}
MZ-TH/00-41\\
CLNS 00/1693\\
hep-ph/0009218
\end{flushright}
\begin{center}
{\Large\bf Analytical calculation of
heavy baryon correlators in NLO of perturbative QCD} \\[1truecm]
{\large S.~Groote,$^{1,2}$ J.G.~K\"orner$^1$ and
  A.A.~Pivovarov$^{1,3}$}\\[.7cm]
$^1$ Institut f\"ur Physik der Johannes-Gutenberg-Universit\"at,\\
  Staudinger Weg 7, 55099 Mainz, Germany\\[.5truecm]
$^2$ Floyd R.~Newman Laboratory, Cornell University, Ithaca, NY 14853,
USA\\[.5truecm]
$^3$ Institute for Nuclear Research of the\\
Russian Academy of Sciences, Moscow 117312, Russia
\vspace{1truecm}
\end{center}

\begin{abstract}
We present analytical next-to-leading order results for the correlator of 
baryonic currents at the three-loop level with one finite mass quark. We
obtain the massless and the HQET limits of the correlator as particular cases
from the general formula, we also give explicit expressions for the moments
of the spectral density. Calculations have been performed with an extensive
use of the symbolic manipulation programs MATHEMATICA and REDUCE.
\end{abstract}

\vskip 1cm 
\begin{center}
{\it Prepared for the ``VII International Workshop on Advanced
Computing\\  
and  Analysis  Techniques in  Physics Research (ACAT2000)''}
\end{center}
\newpage
Baryons form a rich family of particles which has been experimentally studied
with high accuracy~\cite{PDG}. A theoretical analysis of these experimental
data gives a lot of information about the structure of QCD and the numerical
values of its parameters. The hypothetical limit $N_c\to\infty$ for the number
$N_c$ of colours which is a very powerful tool for investigating the general
properties of gauge interactions was especially successful for
baryons~\cite{NcInfty}. The spectrum of baryons is contained in the correlator
of two baryonic currents and the spectral density associated with it. To
leading order the correlator is given by a product of $N_c$ fermionic
propagators. The diagrams of this topology have recently been studied in
detail~\cite{sun0,sun1,wm1,wm2,wm3,thresh}. They are rather frequently used
in phenomenological applications~\cite{sunapp}. With the advent of new
accelerators and detectors many properties of baryons containing a heavy quark
have been experimentally measured in recent years~\cite{PDG}. However,
theoretical calculations beyond the leading order have not been done for many
interesting cases. In this note we fill up this gap.

We report on the results of calculating the $\alpha_s$ corrections to the
correlator of two baryonic currents with one finite mass quark and two
massless quarks. We give analytical results and discuss the magnitude of the
$\alpha_s$ corrections. The massless and HQET limits are obtained as special
cases. We also present analytical results for the moments of the spectral
density associated with the correlator. The extensive discussion of the
impact of our new result for the correlator on the phenomenology of baryons
will be given elsewhere. Note that the massless case has been known since long
ago~\cite{pivbar}. The mesonic analogue of our baryonic calculation was
completed some time ago~\cite{Generalis} and has subsequently provided a rich
source of inspiration for many applications in meson physics. 

\begin{figure}[htb]\begin{center}
\vbox{
\epsfig{figure=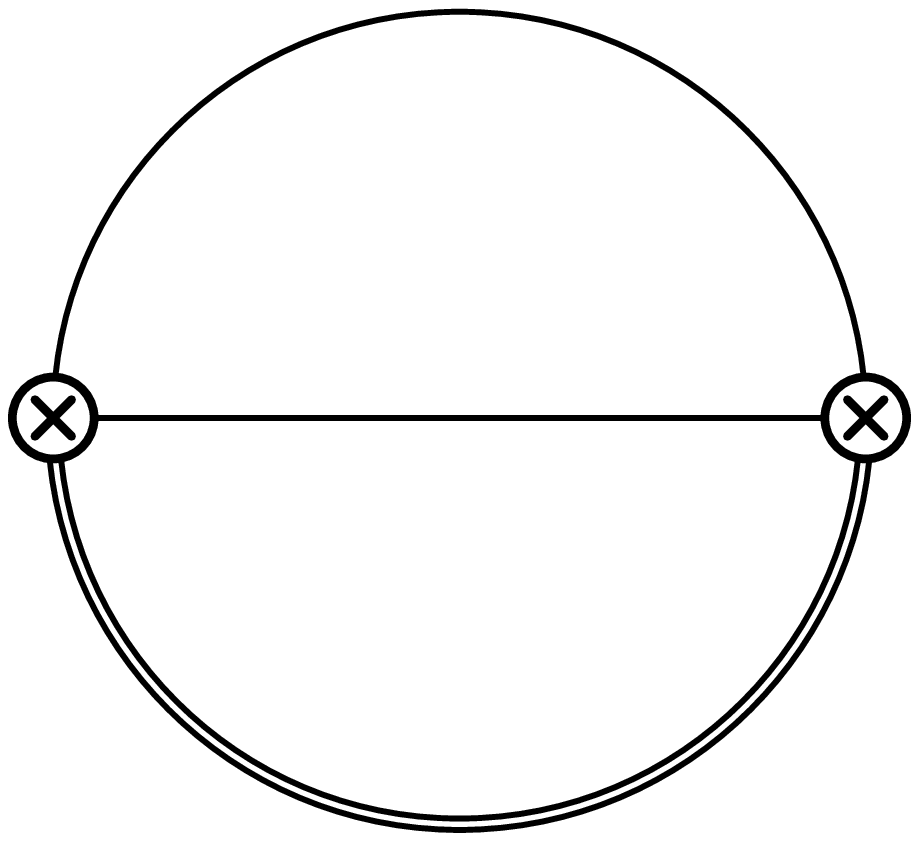, scale=0.3}
\epsfig{figure=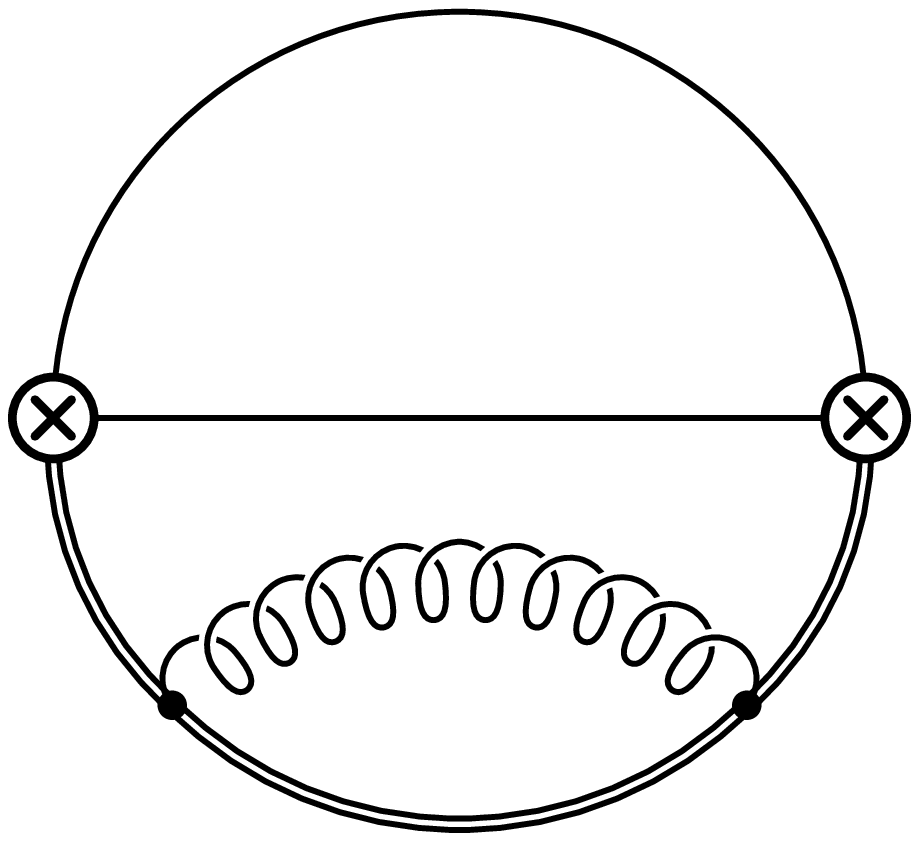, scale=0.3}
\epsfig{figure=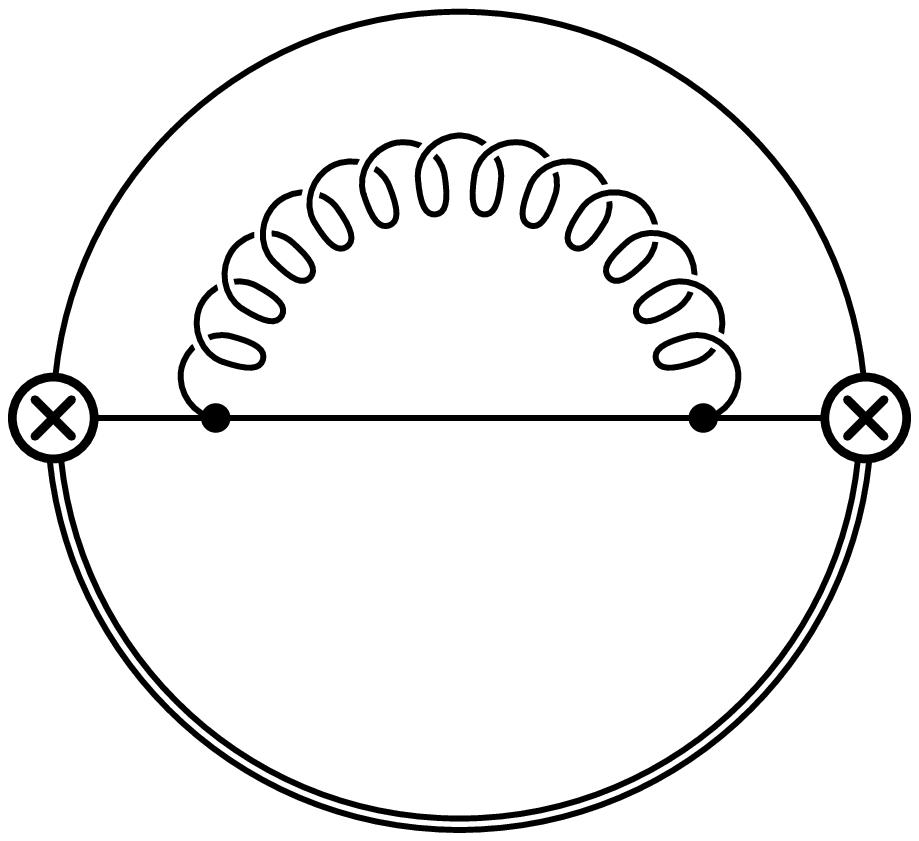, scale=0.3}
\epsfig{figure=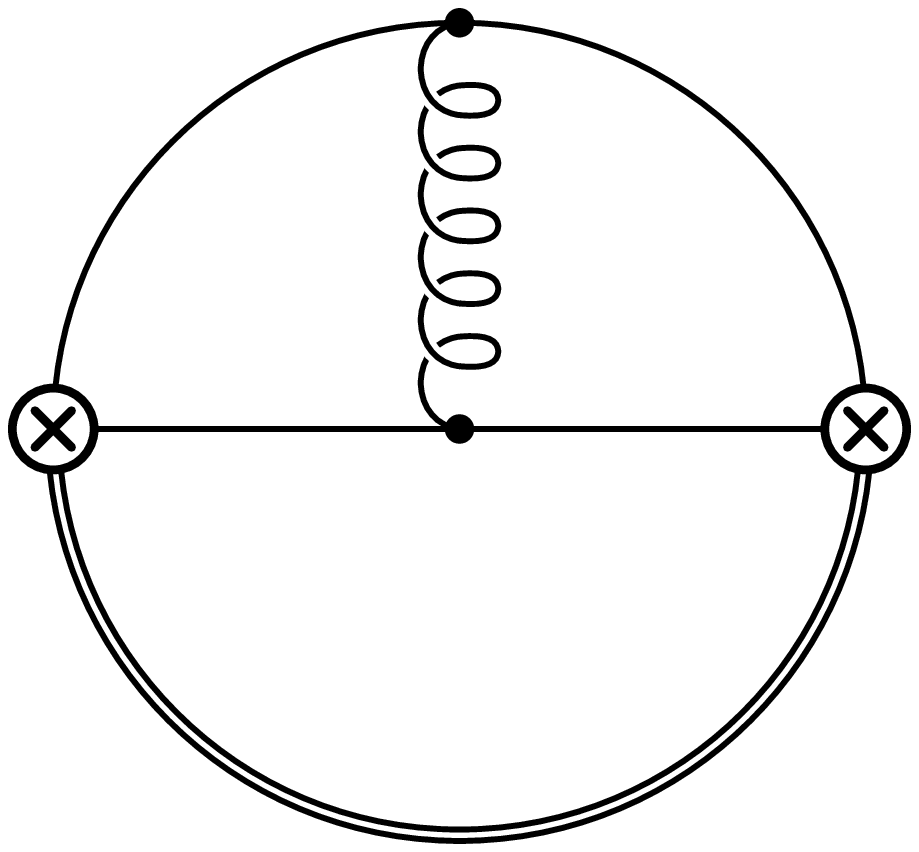, scale=0.3}
\epsfig{figure=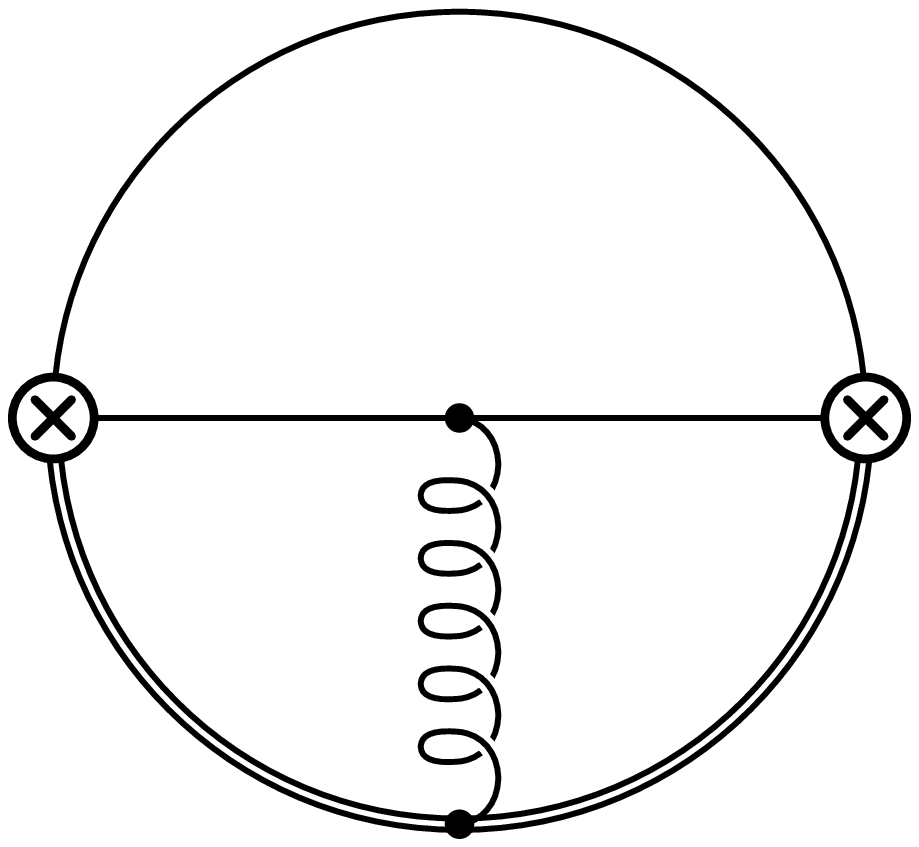, scale=0.3}\\
  (a)\kern68pt(b)\kern68pt(c)\kern68pt(d)\kern68pt(e)}
\caption{\label{fig1}The calculated (a) two-loop and
  (b--e) three-loop topologies}\end{center}
\end{figure}

A generic baryonic current has the form
\begin{equation}\label{cur}
j=\epsilon^{abc}(u_a^T C d_b)\Gamma\Psi_c
\end{equation}
which has the quantum numbers of a $J^P=1/2^-$ baryon for $\Gamma=\gamma_5$.
$\Psi$ is a finite mass quark field with the mass parameter $m$, $u$ and $d$
are massless quark fields, $C$ is the charge conjugation matrix,
$\epsilon^{abc}$ is the totally antisymmetric tensor and $a,b,c$ are colour
indices for the $SU(3)$ colour group. Other baryonic currents with any given
specified quantum numbers are obtained from the current in Eq.~(\ref{cur}) by
inserting the appropriate Dirac matrices. Such additions introduce no
principal complication into our method of calculation. In the following we
take $\Gamma=1$. The correlator of two baryonic currents is expanded as
\begin{equation}\label{def00}
i\int\langle Tj(x)\bar j(0)\rangle e^{iqx}dx
  =\gamma_\nu q^\nu\Pi_q(q^2)+m\Pi_m(q^2).
\end{equation}
$\Gamma=\gamma_5$ leads to the trivial change $\Pi_q(q^2)\to -\Pi_q(q^2)$.
The result for the invariant function $\Pi_m(q^2)$ has already been presented
in Ref.~\cite{masspart}. In this note we show results for the function
$\Pi_q(q^2)$ and compare it with $\Pi_m(q^2)$. The invariant functions
$\Pi_{q,m}(q^2)$ can be represented compactly via the dispersion relation 
\begin{equation}
\Pi_\#(q^2)=\frac1{128\pi^4}\int_{m^2}^\infty\frac{\rho^\#(s)ds}{s-q^2}
\end{equation}
where $\rho^\#(s)=\rho^{q,m}(s)$ are the spectral densities. All quantities
are understood to be appropriately regularized. The spectral density is the
real object of interest for phenomenological applications and we limit our
subsequent discussion therefore to the spectral density
\begin{equation}
\rho^\#(s)=s^2\left\{\rho_0^\#(s)\left(1+\frac{\alpha_s}\pi
  \ln\pfrac{\mu^2}{m^2}\right)+\frac{\alpha_s}\pi\rho_1^\#(s)\right\}.
\end{equation}
Here $\mu$ is the renormalization scale parameter, $m$ is a pole mass of the
heavy quark (see e.g.\ Ref.~\cite{polemass}) and $\alpha_s=\alpha_s(\mu)$.
The leading order two-loop contribution is shown in Fig.~\ref{fig1}(a).
Note that this topology coincides with water melon diagrams for which 
a general method of calculation (with arbitrary masses) has recently been
developed~\cite{wm1,wm2,wm3}. The leading order results read
\begin{equation}\label{lead0q}
\rho_0^q(s)=\frac14-2z+2z^3-\frac14z^4-3z^2\ln z,
\end{equation}
\begin{equation}\label{lead0m}
\rho_0^m(s)=1+9z-9z^2-z^3+6z(1+z)\ln z 
\end{equation}
with $z=m^2/s$. The next-to-leading order contribution is given by three-loop
diagrams with one external momentum. For an arbitrary mass arrangement such
diagrams have not yet been calculated analytically. However, if we take the
case of one massive line, the result within $\msbar$-scheme can be obtained
analytically and reads
\begin{eqnarray}\label{corr1q}
\lefteqn{\rho_1^q(s)\ =\ \frac{71}{48}-\frac{565}{36}z-\frac78z^2
  +\frac{625}{36}z^3-\frac{109}{48}z^4}\\&&\kern-14pt
  -\left(\frac{49}{36}-\frac{116}9z+\frac{116}9z^3-\frac{49}{36}z^4\right)
  \ln(1-z)+\left(\frac14-\frac{17}3z-11z^2+\frac{113}9z^3-\frac{49}{36}z^4
  \right)\ln z\nonumber\\&&\kern-14pt
  +\left(\frac13-\frac83z+\frac83z^3-\frac13z^4\right)\ln(1-z)\ln z
  -2z^2\left(9+\frac43z-\frac16z^2\right)\left(\frac12\ln^2z-\zeta(2)\right)
  \nonumber\\&&\kern-14pt
  +\left(\frac23-\frac{16}3z-18z^2+\frac83z^3-\frac13z^4\right)\Li_2(z)
  -12z^2\left(\Li_3(z)-\zeta(3)-\frac13\Li_2(z)\ln(z)\right)\nonumber
\end{eqnarray}
where $\Li_n(z)$ are polylogarithms and $\zeta(n)$ is Riemann's zeta function,
$\zeta(2)=\pi^2/6$. The contributing three-loop diagrams are shown in
Figs.~\ref{fig1}(b) to (e). They have been evaluated using advanced algebraic
methods for multi-loop calculations along the lines decribed in
Refs.~\cite{wm2,Generalis}. This result should be compared to 
\begin{eqnarray}\label{corr1m}
\lefteqn{\rho_1^m(s)\ =\ 9+\frac{665}9z-\frac{665}9z^2-9z^3}\\&&
  -\left(\frac{58}9+42z-42z^2-\frac{58}9z^3\right)\ln(1-z)
  +\left(2+\frac{154}3z-\frac{22}3z^2-\frac{58}9z^3\right)\ln z\nonumber\\&&
  +4\left(\frac13+3z-3z^2-\frac13z^3\right)\ln(1-z)\ln z
  +12z\left(2+3z+\frac19z^2\right)\left(\frac12\ln^2z-\zeta(2)\right)
  \nonumber\\&&
  +4\left(\frac23+12z+3z^2-\frac13z^3\right)\Li_2(z)
  +24z(1+z)\left(\Li_3(z)-\zeta(3)-\frac13\Li_2(z)\ln z\right).\nonumber
\end{eqnarray}
Our method of integration is a completely algebraic one and therefore symbolic
manipulation programs can be used for performing the long calculations. Two
independent calculations of some steps were done using REDUCE and MATHEMATICA.
REDUCE is rather actively used for high energy calculations (see e.g.\
Ref.~\cite{pivAi}). All diagrams have first been reduced to scalar prototypes.
The integrals over massless loops have been performed (for recurrent
integration where possible) and one is left with the basic integral
\begin{equation}
V(\alpha,\beta;q^2/m^2)=\int\frac{d^Dk}{(k^2+m^2)^\alpha(q-k)^{2\beta}}
\end{equation}
which is a generalization of the standard object $G(\alpha,\beta)$ of the
massless calculation~\cite{Gsch,intbyparts}. The integral $V$ is known
analytically and it suffices in order to calculate the diagrams in
Figs.~\ref{fig1}(c) and~(d). For the calculation of the diagram shown in
Fig.~1(e) the basic integral $V$ enters as a subdiagram. This subdiagram then
is represented in terms of a dispersion integral which makes the whole diagram
computable in terms of the same $V$ with the argument depending on the loop
momentum. The final step is a finite range (convolution type) integration over
this internal momentum with a spectral density of the basic integral $V$. The
reduction to scalar prototypes of the diagram shown in Fig.~\ref{fig1}(e)
leads also to a new irreducible block (i.e.\ a prototype not expressible in
terms of $V$) which is related to a two-loop master (fish) diagram. The result
for this diagram is taken from Ref.~\cite{Broad}.

The results given in Eqs.~(\ref{corr1q}) and~(\ref{corr1m}) represent the full
next-to-leading order solution. Since the anomalous dimension of the current
in Eq.~(\ref{cur}) is known up to two-loop order~\cite{pivan}, the results
shown in Eqs.~(\ref{corr1q}) and~(\ref{corr1m}) complete the ingredients
necessary for an analysis of the correlator in Eq.~(\ref{def00}) within
operator product expansion at the next-to-leading order level. 

We now turn to the analysis of Eq.~(\ref{corr1q}) and contrast it with the
corresponding results for $\rho^m(s)$ as published in Ref.~\cite{masspart}.
Two limiting cases of general interest are the near-threshold and the high
energy asymptotics. With our result given in Eq.~(\ref{corr1q}) both limits
can be taken explicitly. The asymptotic expressions can be also obtained in
the framework of effective theories which can be viewed as special devices for
such calculations.

In the high energy (or, equivalently, small mass) limit $z\rightarrow 0$ the
corrections read
\begin{eqnarray}\label{masslessq}
\rho_1^q(s)&=&\frac{71}{48}+\frac14\ln z-\frac{41}3z-6z\ln z+O(z^2),\\
\label{masslessm}
\rho_1^m(s)&=&9+83z-4\pi^2z+2\ln z+50z\ln z+12z\ln^2z-24z\zeta(3)+O(z^2).\qquad
\end{eqnarray}
Therefore we obtain
\begin{eqnarray}
\lefteqn{\rho^q(s)\ =\ \rho_{\rm massless}^q(s)}\label{q00}\\
  &=&\frac{s^2}4\left\{1+\frac{\alpha_s}\pi
  \left(\ln\pfrac{\mu^2}s+\frac{71}{12}\right)\right\}
  -2m_{\msbar}^2(\mu)s\left\{1+\frac{\alpha_s}\pi
  \left(3\ln\pfrac{\mu^2}s+\frac{19}2\right)\right\},\nonumber\\
\lefteqn{m\rho^m(s)\ =\ m_{\msbar}(\mu)\rho_{\rm massless}^m(s)
  \ =\ m_{\msbar}(\mu)s^2\left\{1+\frac{\alpha_s}\pi
  \left(2\ln\pfrac{\mu^2}s+\frac{31}3\right)\right\}}\label{m00}
\end{eqnarray}
where $\rho_{\rm massless}^\#(s)$ is the result of calculating the correlator
in the effective theory of massless quarks. For the momentum part $\rho^q(s)$
we retain the $O(m^2)$ correction. The relation between the pole mass $m$ and
the $\msbar$ mass $m_{\msbar}(\mu)$ we have used reads
\begin{equation}
m=m_{\msbar}(\mu)\left\{1+\frac{\alpha_s}\pi
  \left(\ln\pfrac{\mu^2}{m^2}+\frac43\right)\right\}.
\end{equation}
Note that the massless effective theory cannot reproduce the mass
singularities (terms like $z\ln(z)$ in Eq.~(\ref{masslessm})). These 
singularities can be parametrized by condensates of local operators.
The first $m^2$ correction in Eqs.~(\ref{lead0m}) and~(\ref{masslessm})
(or, equivalently, the $m^3$ correction to the expression in Eq.~(\ref{m00})) 
can be found if the perturbative value of the heavy quark condensate
$\langle 0|\bar\Psi\Psi|0\rangle$ taken from the full theory is
added~\cite{polit}. The composite operator $(\bar\Psi\Psi)$ should
be understood within a mass independent renormalization scheme such as the
$\msbar$-scheme. This value (perturbatively,
$\langle 0|\bar\Psi\Psi|0\rangle\sim m^3\ln(\mu^2/m^2)$) cannot be computed
within the effective theory of massless quarks. It provides the proper matching
between the perturbative expressions for the correlators of the full (massive)
and effective (massless) theories. This matching procedure allows one to
restore higher order terms of the mass expansion in the full theory from the
effective massless theory with the mass term treated as a
perturbation~\cite{CheSpi88}. The account for the mass term as a perturbation
in a massless theory is justified at high energies and greatly simplifies the
calculations (see e.g.\ Ref.~\cite{kwia}). Note that the correction of order
$m^2/s$ to $\rho^m(s)$ can actually be found in this manner because it depends
only on one local operator $(\bar\Psi\Psi)$ and, therefore, the calculation is
technically feasible. In case of the function $\rho^q(s)$ the situation is
different because there is no gauge invariant operator of dimension two in the
effective massless theory. Therefore, the mass singularities of the form
$m^2\log(m^2/s)$ should not appear in the expansion for $\rho^q(s)$ at large
energies. The result in Eq.~(\ref{q00}) shows this explicitly. Note that the
absence of such singularities is one of our checks of the correctness of the
calculation.

In the near-threshold limit $E\rightarrow 0$ with $s=(m+E)^2$ one explicitly
obtains
\begin{eqnarray}\label{hqet0}
\rho_{\rm thr}^m(m,E)&=&\frac{16E^5}{5m}
  \Bigg\{1+\frac{\alpha_s}\pi\ln\pfrac{\mu^2}{m^2}\\&&\qquad\qquad
  +\frac{\alpha_s}\pi\left(\frac{54}5+\frac{4\pi^2}9+4\ln\pfrac{m}{2E}\right)
  \Bigg\}+O\pfrac{E^6}{m^2}.\nonumber
\end{eqnarray}
The invariant function $\rho^m(s)$ suffices to determine the complete leading
HQET behaviour since one has $\slq\rho^q+m\rho^m\rightarrow
(\slv+1)\rho_{\rm HQET}$ for the leading term. We explictly checked this
relation. In this region the appropriate device to compute the limit of the
correlator is HQET (see e.g.\ Refs.~\cite{Geor,Neub}). Writing
\begin{equation}\label{hqet}
m\rho_{\rm thr}^m(m,E)=C(m/\mu,\alpha_s)^2\rho_{\rm HQET}(E,\mu)
\end{equation}
we obtain the known result for $\rho_{\rm HQET}(E,\mu)$~\cite{groote} 
\begin{equation}\label{explrhoHQET}
\rho_{\rm HQET}(E,\mu)=\frac{16E^5}5\left\{1+\frac{\alpha_s}\pi
  \left(\frac{182}{15}+\frac{4\pi^2}9+4\ln\frac{\mu}{2E}\right)\right\}+O(E^6)
\end{equation}
with the matching coefficient $C(m/\mu,\alpha_s)$ given by~\cite{barmatch}
\begin{equation}\label{matchcoef}
C(m/\mu,\alpha_s)
  =1+\frac{\alpha_s}\pi\left(\frac12\ln\pfrac{m^2}{\mu^2}-\frac23\right).
\end{equation}
The matching procedure allows one to restore the near-threshold limit of the
full correlator starting from the simpler effective theory near the
threshold~\cite{Ein}.

Note that the higher order corrections in $E/m$ to Eq.~(\ref{hqet0}) can
easily be obtained from the explicit result given in Eq.~(\ref{corr1q}). 
Indeed, the next-to-leading order corrections in low energy expansion read
\begin{eqnarray}\label{threq}
\Delta\rho_{\rm thr}^q(m,E)&=&-\frac{8E^6}{m^2}
  \Bigg\{1+\frac{\alpha_s}\pi\left(\ln\pfrac{\mu^2}{m^2}
  +\frac{908}{75}+\frac{4\pi^2}9+\frac{68}{15}\ln\pfrac{m}{2E}\right)\Bigg\},\\
\Delta\rho_{\rm thr}^m(m,E)&=&-\frac{24E^6}{5m^2}\left\{1+\frac{\alpha_s}\pi
  \left(\ln\pfrac{\mu^2}{m^2}+\frac{584}{45}+\frac{4\pi^2}9
  +\frac{44}9\ln\pfrac{m}{2E}\right)\right\}.\qquad\label{threm}
\end{eqnarray}
To obtain this result starting from HQET is a more difficult task requiring
the analysis of contributions of higher dimension operators. 

\begin{figure}[htb]\begin{center}
\epsfig{figure=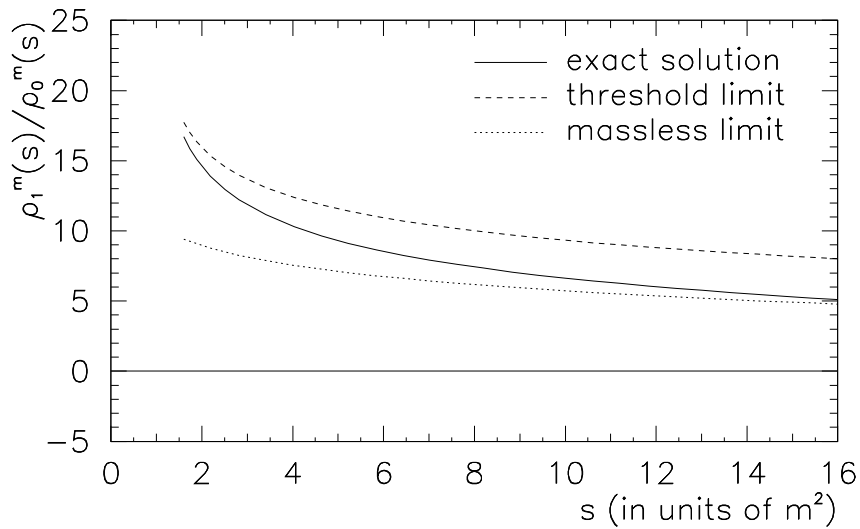, scale=0.7}
\caption{\label{fig2}The ratio $\rho_1^m(s)/\rho_0^m(s)$ of the next-to-leading
correction and the leading order term in dependence of the energy square $s$}
\end{center}\end{figure}
We now discuss some quantitative features of the correction given in
Eq.~(\ref{corr1q}). Of interest is whether the two limiting expressions
(the massless limit expression as given in Eq.~(\ref{q00}) and the HQET limit
expression in Eqs.~(\ref{hqet0}) and~(\ref{hqet})) can be used to characterise 
the full function for all energies.

\begin{figure}[htb]\begin{center}
\epsfig{figure=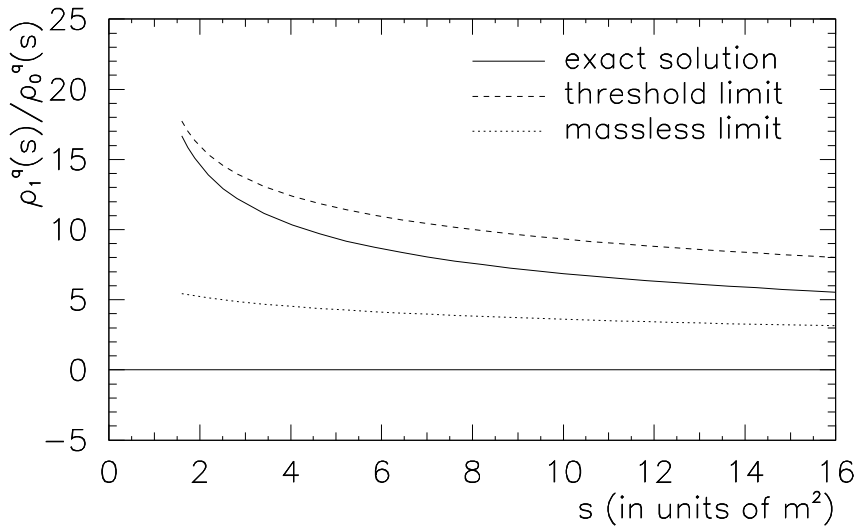, scale=0.7}
\caption{\label{fig3}The ratio $\rho_1^q(s)/\rho_0^q(s)$ of the next-to-leading
correction and the leading order term in dependence of the energy square $s$}
\end{center}\end{figure}

For this discussion we compare components of the baryonic spectral function in
leading and next-to-leading order. In Fig.~\ref{fig2} and~\ref{fig3} we show
the ratio $\rho_1^\#(s)/\rho_0^\#(s)$ for $\#=m$ and $\#=q$, respectively. In
the following we shall always use the specific renormalization scale value
$\mu=m$ if it is not written explicitly.
One can see that a simple interpolation between the two limits can give a
rather good approximation for the next-to-leading order correction in the
complete region of $s$.

The informative set of observables are moments of the spectral density
\begin{equation}
{\cal M}_n^\#=\int_{m^2}^\infty\frac{\rho^\#(s)ds}{s^n}=m^{6-2n}M_n^\#
\end{equation}
where $M_n^\#$ are dimensionless quantities. We find 
\begin{equation}\label{resform1}
M_n^\#=M_n^{\#{(0)}}\left\{1+\frac{\alpha_s}\pi
  \left(\ln\pfrac{\mu^2}{m^2}+\delta_n^\#\right)\right\}
\end{equation}
where 
\begin{eqnarray}
M_n^{q(0)}&=&\frac{12}{(n+1)n(n-1)^2(n-2)(n-3)},\\
M_n^{m(0)}&=&\frac{12}{n(n-1)^2(n-2)^2(n-3)},
\end{eqnarray}
and
\begin{equation}
\delta_n^\#=A_n^\#+\frac{2\pi^2}9.
\end{equation}
The coefficients $A_n^\#$ are rational numbers, and the closed form expressions
for $\delta_n^\#$ are long. Therefore we only show the first values for $A_n^q$
in the second column of Table~\ref{tab1}. They are numerically rather close to
the results obtained for $A_n^m$ in Ref.~\cite{masspart}.
\begin{table}[htb]\begin{center}
\begin{tabular}{|r||c|l|}\hline
$n$&$A^q_n$&$\delta^q_n-\delta^q_{n-1}$\\\hline
$4$&$9/2$&\\
$5$&$22/3$&$2.833333$\\
$6$&$109/12$&$1.750000$\\
$7$&$5593/540$&$1.274074$\\
$8$&$6133/540$&$1.000000$\\
$9$&$460351/37800$&$0.821190$\\
$10$&$40553/3150$&$0.695370$\\
$11$&$148574/11025$&$0.602132$\\
$12$&$2470739/176400$&$0.530357$\\
$13$&$758614613/52390800$&$0.473463$\\
$14$&$156200257/10478160$&$0.427302$\\
$15$&$4583939335/299675376$&$0.389128$\\
$16$&$117273501721/7491884400$&$0.357055$\\
$17$&$3113341968041/194788994400$&$0.329746$\\
$18$&$3172990981751/194788994400$&$0.306224$\\
$19$&$54887116886639/3311412904800$&$0.285760$\\
$20$&$111547839702373/6622825809600$&$0.267802$\\
$21$&$7313770708819951/427834547300160$&$0.251920$\\
$22$&$1483100149208267/85566909460032$&$0.237779$\\
$23$&$142724395992842749/8128856398703040$&$0.225109$\\
\hline\end{tabular}
\caption{\label{tab1}Values for the rational part $A^q_n$ of the first moments
  $\delta^q_n$ and their relative difference $\delta^q_n-\delta^q_{n-1}$}
\end{center}\end{table}

By representing the moments through
\begin{equation}\label{resform1norm}
\frac{M_n^\#}{M_n^{\#(0)}}=\frac{M_N^\#}{M_N^{\#(0)}}
\left\{1+\frac{\alpha_s}\pi(\delta_n^\#-\delta_N^\#)\right\}
\end{equation}
we have all corrections to be normalized to the moment $M_N^\#$ of fixed order
$N$. Note that the difference $\delta_n^\#-\delta_N^\#$ is scheme-independent.
This feature can be used in the high precision analysis of heavy quark
properties~\cite{bbmass} within NRQCD (see e.g.\ Ref.~\cite{hoang}). With
Eq.~(\ref{resform1norm}) one can find the actual (invariant or
scheme-independent) magnitude of the correction. Indeed, for any given $N$ one
can find a set of perturbatively commensurate moments $M_n^\#$ with $n\sim N$
for which the requirement of the chosen precision is satisfied. In the third
column of Table~\ref{tab1} we therefore present numerical values only for the
differences of the $\delta_n^q$ for subsequent orders.

Note that the moments represent massive vacuum bubbles, i.e.\ diagrams with
massive lines without external momenta. These diagrams have been
comprehensively analyzed in Refs.~\cite{avdeev,david}. The analytical results
for the first few moments at three-loop level can be checked independently
with existing computer programs for symbolic calculations in high energy
physics (see e.g.\ Ref.~\cite{ChKStmom8}).

The presented results have phenomenological applications within the sum rule
analysis of baryon properties (see e.g.\ Refs.~\cite{ioffe,chung,zphys}).
As an example we calculate the integral of $\rho^q(s)$ up to the energy
cut $\sqrt{s_0}$,
\begin{equation}\label{residue}
{\cal M}_0^q(s_0)=\int_{m^2}^{s_0}\rho^q(s)ds
\end{equation}
which is related to the coupling constant (residue) of a baryon to the
current in Eq.~(\ref{cur}). In NLO the integral is represented by 
\begin{equation}\label{residueBreak}
{\cal M}_0^q(s_0)={\cal M}_0^{q(0)}(s_0)\left(1+\frac{\alpha_s}\pi
\left(\ln\pfrac{\mu^2}{m^2}+\Delta(s_0)\right)\right)
\end{equation}
which leads to the renormalization of the LO result for the residue in
the form
\begin{equation}\label{residueRenorm}
Z_R^{01}=\frac{{\cal M}_0^q(s_0)}{{\cal M}_0^{q(0)}(s_0)}
  =1+\frac{\alpha_s}\pi\left(\ln\pfrac{\mu^2}{m^2}+\Delta(s_0)\right)
  +O(\alpha_s^2).
\end{equation}
For e.g.\ $s_0=2m^2$, $\mu^2=m^2$ we find numerically
\begin{equation}\label{residueRenormNum}
Z_R^{01}=1+\frac{\alpha_s}\pi\Delta(2m^2)=1+\frac{\alpha_s}\pi 15.4117\ldots
\end{equation}
We see that the NLO correction to the residue in the $\msbar$-scheme is rather
large. For the numerical value of the coupling constant $\alpha_s\approx 0.3$
which is a typical value for baryons containing a $c$-quark, the NLO
correction in the $\msbar$-scheme reaches the $100\%$ level. 

One can see that the corrections to the moments basically reflect the shape of
the correction to the spectrum. Even the massless approximation is reasonably
good for relative corrections for the first few moments despite of the
unfavorable shape of the weight function $1/s^n$. It can be improved by
changing the subtraction point $\mu$, i.e.\ by switching from the
$\msbar$-scheme to some other renormalization scheme, or by resumming the
integrand~\cite{pivtau} which lies beyond the scope of finite order
perturbation theory though.

To conclude, we have computed the next-to-leading perturbative corrections to
the finite mass baryon correlator at three-loop order. Technically, the method
allows one to obtain analytical results for two-point correlators of composite
operators with one finite mass particle that can be compared to HQET results.
Corrections in $E/m$ near threshold are easily available from our explicit
results. From threshold to high energies the exact spectral density
interpolates nicely between the leading order HQET result close to threshold
and the asymptotic mass zero result. Going even one order higher it is very
likely that the full four-loop spectral density can be well approximated by
the corresponding massless four-loop result which can be calculated using
existing computational algorithms~\cite{ibyparts,CheSmi}.

{\bf Acknowledgements:}
The present work is supported in part by the Volkswagen Foundation under
contract No.~I/73611 and by the Russian Fund for Basic Research under contract
99-01-00091. A.A.~Pivovarov is an Alexander von Humboldt fellow. S.~Groote
gratefully acknowledges a grant given by the DFG, FRG.

\end{document}